\documentclass[aps,reprint,twocolumn,superscriptaddress,citeautoscript,showpacs,amsmath,amssymb,floatfix,prb,longbibliography]{revtex4-1}
\usepackage{}%\bibliographystyle{apsrev}
\usepackage{graphicx}
\usepackage{amsmath}
\usepackage{amssymb}
\usepackage{mathtools}
\usepackage{textcomp,gensymb} %For degree symbol
\usepackage{bm} %For bold math (Greek) symbols or bold italic in math mode
\usepackage[usenames,dvipsnames]{xcolor}
\usepackage{float}
\usepackage{setspace}
\usepackage{dcolumn}
\usepackage{array}

\usepackage{changepage}
\usepackage{tikz}
\usepackage{contour}
\usepackage[left]{lineno}
\usepackage{blindtext}
\usepackage{setspace}
\usepackage[none]{hyphenat}%for removing hyphen
\usepackage[colorlinks=true,citecolor=blue,linkcolor=blue,pdftex,hypertexnames=false,urlcolor=blue,linktocpage=true]{hyperref}%for hyperlink
\definecolor{ashgray}{rgb}{0.7,0.75,0.71}
\definecolor{mspringgreen}{rgb}{0, 0.8, 0.1}
\definecolor{auburn}{rgb}{0.43, 0.21, 0.1}
\definecolor{ao(english)}{rgb}{0.0, 0.5, 0.0}
\definecolor{afw}{rgb}{0.95, 0.95, 0.96}
\definecolor{magnolia}{rgb}{0.97, 0.96, 1.0}
\definecolor{wsmk}{rgb}{0.96, 0.96, 0.96}

\newcolumntype{M}[1]{>{\centering\arraybackslash}m{#1}}

\usepackage{dcolumn}% Align table columns on decimal point
\usepackage{multirow} 
\usepackage{soul}
\usepackage{titlesec}
\usepackage{color}

\usepackage{bbold}

\newcolumntype{d}[1]{D{.}{.}{#1}}

\begin{document}

\title{Room-temperature shape-memory effect in Sr(Ni$_{1-x}$Cu$_x$)$_2$P$_2$}

\author{Juan Schmidt}
%\altaffiliation{Currently at Instituto de F\'isica de Buenos Aires, CONICET-Universidad de Buenos Aires, Pabell\'on 1, Ciudad Universitaria, CABA, 1428, Argentina.}
\affiliation{Department of Physics and Astronomy, Iowa State University, Ames, IA 50011, USA}
\affiliation{Ames National Laboratory, Iowa State University, Ames, IA 50011, USA}
\affiliation{IFIBA, CONICET, Pabell\'on 1, Ciudad Universitaria, C.A.B.A, 1428,
Argentina.}
\author{Alexander Horvath}
\affiliation{Department of Materials Science and Engineering and Institute of Materials Science, University of Connecticut, 97 North Eagleville Road, Unit 3136, Storrs, Connecticut 06269-3136, USA}
\author{Seok-Woo Lee}
\affiliation{Department of Materials Science and Engineering and Institute of Materials Science, University of Connecticut, 97 North Eagleville Road, Unit 3136, Storrs, Connecticut 06269-3136, USA}
\author{Sergey L. Bud'ko}
\affiliation{Department of Physics and Astronomy, Iowa State University, Ames, IA 50011, USA}
\affiliation{Ames National Laboratory, Iowa State University, Ames, IA 50011, USA}
\author{Paul C. Canfield}
\affiliation{Department of Physics and Astronomy, Iowa State University, Ames, IA 50011, USA}
\affiliation{Ames National Laboratory, Iowa State University, Ames, IA 50011, USA}

\date{\today}

\pacs{1234}

\begin{abstract}

The compound SrNi$_2$P$_2$ can exhibit multiple crystal structures with no P-P pairs bonded (uncollapsed tetragonal, or ucT, state), with one-third of the P-P pairs bonded (one-third collapsed orthorhombic, or tcO, state), or with all P-P pairs bonded (collapsed tetragonal, or cT, state) across the Sr layers. The system can be tuned into its different states by changing temperature, mechanical stress, or chemical composition. Changes in bonding may manifest in changes of macroscopic properties of the material, such as its shape, electrical conductivity, or magnetism. In this work, we show that SrNi$_2$P$_2$ can be tuned among the three states by changing Cu substitution and temperature. We present temperature-dependent resistance and single-crystal x-ray diffraction results in Sr(Ni$_{1-x}$Cu$_x$)$_2$P$_2$ single-crystals that show that Cu substitution favors the P-P bonding, stabilizing the cT state at ambient pressure. We construct a $T-x$ phase diagram that shows how all of these transition temperatures increase with increasing Cu fraction, $x$. The transition between the tcO state and the cT state exhibits a very large thermal hysteresis, which can be tuned to temperatures close to room temperature. In particular, the properties of Sr(Ni$_{0.963}$Cu$_{0.037}$)$_2$P$_2$ may make it suitable for applications as a shape memory material at room temperature.

\end{abstract}

\maketitle
\section{Introduction}

Compounds with ThCr$_2$Si$_2$ structural motifs are not only interesting for their diverse electronic and magnetic properties, but also for the remarkable mechanical properties that many of them exhibit. This structure can be adopted by hundreds of compounds with formula $AM_2X_2$ with $A$ usually being an alkali, alkaline-earth, or rare earth element, $M$ being a transition metal and $X$ being a representative element of the $p$-block \cite{Shatruk2019}. Their structure can be classified mainly as \textit{collapsed tetragonal} (cT) \cite{Kreyssig2008} when the $X$-$X$ pairs bond across the $A$ layers, or \textit{uncollapsed tetragonal} (ucT) when they do not. Other compounds with related structures with formula $ABM_4X_4$ can not only exhibit the cT structure when the $X$ atoms bond across all the alternating $A$ and $B$ layers, but also an intermediate state known as a \textit{half-collapsed tetragonal structure} (hcT) when they bond only across the $A$ layer and not the $B$ layer \cite{Kaluarachchi2017, Borisov2018,Xiang2018b,Xiang2022,Stillwell2019,Wang2023,Song2021}. Additionally, SrNi$_2$P$_2$ can adopt another structure, known as \textit{one-third collapsed orthorhombic} (tcO) \cite{Schmidt2023}, where only one third of the P atoms bond across the Sr layers, resulting in an orthorhombic supercell \cite{Keimes1997}.

The ability of some of these materials to reversibly transition from one structure to another upon applying hydrostatic pressure \cite{Kreyssig2008,Gati2012,Kaluarachchi2017,Borisov2018,Torikachvili2008,Yu2009,Huyan2025} or anisotropic strains \cite{Xiao2021,Sypek2017} provides these materials with unique mechanical properties that make them potentially interesting for engineering applications \cite{Juan2009,Huber1997}. The maximum recoverable strain quantifies how much a material can change its length without being permanently deformed (or even breaking), and has been shown to reach as high as 17\% in compounds with ThCr$_2$Si$_2$-related structures such as CaFe$_2$As$_2$, CaKFe$_4$As$_4$ and SrNi$_2$P$_2$ \cite{Sypek2017,Song2019,Bakst2018,Xiao2021}, given by the combined contribution of the elastic deformation and the deformation associated with their structural phase transitions, both of which are reversible. Materials with high maximum recoverable strain due to strain-induced reversible structural phase transitions are known as \textit{pseudoelastic} or \textit{superelastic} whenever they recover their original shape by just removing the applied load \cite{Omori2013,Finlayson2023}. In some special cases, the material may retain its deformation after the load is removed, but be able to return to its original shape by heating or applying a magnetic field \cite{Omori2013,Robinson1976}. These materials are known as \textit{shape memory alloys}, which have applications as sensors \cite{Birman1997} and actuators \cite{Huber1997}. One of the main challenges for these applications consists in finding materials that resist multiple mechanical cycles with a minimum fatigue.

SrNi$_2$P$_2$ is an excellent candidate for mechanical applications for its maximum recoverable strain of $\sim$14\% as well as for its superior fatigue resistance, showing negligible change after over 10,000 cycles of a micro-pillar compression \cite{Xiao2021}. At room temperature this material is in the tcO state and can transition to the ucT state upon warming above 325 K \cite{Keimes1997}. Partial substitution of Ni with Co or Rh can suppress this phase transition to lower temperatures favoring the ucT state with no P-P bonding \cite{Schmidt2023,Schmidt2025}. In this study we show that Cu substitution has the opposite effect of favoring the bonding, enhancing the tcO$\leftrightarrow$ucT transition temperature and even stabilizing the cT state. We show that Cu substitution can tune the material properties and potentially make it a shape memory compound at room temperature. 

\section{Experimental Details}

Single crystals of Sr(Ni$_{1-x}$Cu$_x$)$_2$P$_2$ were obtained by the high-temperature solution growth method \cite{Canfield2020}
from a Sn flux \cite{Schmidt2023}. The pure elements were loaded into a $2\ \text{ml}$ alumina fritted Canfield Crucible Set \cite{CanfieldP.C.KongT.KaluarachchiU.S.2016,LSPCeramics}, and sealed under partial atmosphere of argon in a fused silica tube. The starting compositions of Sr$_{1.3}$(Ni$_{1-x}$Cu$_x$)$_2$P$_{2.3}$Sn$_{16}$ were used. The ampoules were placed inside a box
furnace, ramped over 5 hours to $600\ ^{\circ}\text{C}$, held for 4 hours before increasing to $1150\ ^{\circ}\text{C}$, held for 24 hours to make sure the material was fully melted, and finally slowly cooled down to a temperature $T_{dec}$, at which the excess Sn solution was decanted with the aid of a centrifuge \cite{Canfield2020}. The temperature $T_{dec}$ was adjusted differently for each composition in the range 650--950$^{\circ}$C, in order to control the thickness of the grown crystals as was done in Ref. \citenum{Schmidt2023} to prevent composition gradients on large samples. The use of fritted crucible set was crucial for this purpose as it allows us to try different temperature profiles by reusing the decant \cite{CanfieldP.C.KongT.KaluarachchiU.S.2016}.

The Cu concentration levels ($x$) were determined by Energy-Dispersive x-ray Spectroscopy (EDS) quantitative chemical analysis with an EDS detector (Thermo NORAN Microanalysis System, model C10001) attached to a JEOL scanning-electron microscope (SEM). An acceleration voltage of
$22\ \text{kV}$, working distance of $10\ \text{mm}$ and take-off angle of $35 ^{\circ}$ were used for measuring all standards and crystals with unknown composition. A crystal of SrNi$_2$P$_2$ was used as a standard for Sr, Ni and P quantification, and a crystal of GdCu$_2$ was used as a standard for Cu. The spectra were fitted using NIST-DTSA II Microscopium software \cite{Newbury2014}. The composition of each platelike crystal was measured on at least four different positions on the crystal face (perpendicular to $c$), as well as different points across a polished edge of the crystal (along $c$). The average compositions and error bars were obtained from these data, accounting for both inhomogeneity and goodness of fit of each spectrum. It should be noted that, unlike the Sr(Ni$_{1-x}$Co$_x$)$_2$P$_2$ system \cite{Schmidt2023}, the Sr(Ni$_{1-x}$Cu$_x$)$_2$P$_2$ system \textit{does not} suffer from strong inhomogeneity along the $c$-axis. 

Single-crystal x-ray diffraction was performed using a
Rigaku XtaLab Synergy-S diffractometer with Ag radiation ($\lambda=0.56087\ \text{\AA}$), in transmission mode, operating at 65 kV and 0.67 mA. The samples were held in a nylon loop with Apiezon N grease. The temperature was controlled using Oxford Cryostream 1000, by adjusting the flow of cold nitrogen gas on the crystal. The transition temperatures extracted from the single-crystal x-ray diffraction measurements collected with this instrument for other systems \cite{Schmidt2025} are consistent, within 3 K, with those extracted from other measurement techniques (e.g. temperature-dependent thermodynamic and transport measurements). The total number of runs and images was based on the strategy calculation from the program CrysAlisPro (Rigaku OD, 2023). The data integration and reduction were also performed using CrysAlisPro, and a numerical absorption correction was applied based on Gaussian integration over a face-indexed crystal. The structures were solved by intrinsic phasing using the SHELXT software package and were refined with SHELXL.

 The temperature-dependent AC resistance of the samples was measured in a Quantum Design Physical Property Measurement System (PPMS) using the AC transport (ACT) option, with a frequency of $17\ \text{Hz}$ and a $3\ \text{mA}$ excitation current. Only in-plane resistance was measured, using a standard four-contact geometry. Electrical contacts with less than $1.5\ \Omega$ resistance were achieved by spot welding $25\ \mu\text{m}$ Pt wire to the samples, followed by adding Epotek H20E silver epoxy, and curing the latter for 1 hour at $120^{\circ}\text{C}$. After the contacts were cured, the sample was quenched rapidly to room temperature by removing them from the furnace.

 DC magnetization measurements were carried out in a Quantum Design Magnetic Property Measurement System (MPMS classic) superconducting quantum interference device (SQUID) magnetometer (operated in the range $1.8\ \text{K}\leq T \leq 350\ \text{K}$, in a field of $H = 10\ \text{kOe}$). Both zero field cooling (ZFC) and field cooling (FC) protocols were used. Each sample was measured with the field applied perpendicular to the tetragonal $c$ axis. The samples were held in between two longitudinal plastic straws inside an outer straw without using any glue in order to reduce strains due to differential thermal contraction \cite{Boehmer2017}, which may affect the collapsed tetragonal transitions. Small amounts of teflon tape were used to cover the bottoms of the straws so that when/if a sample shattered upon cooling through a cT transition, the fine pieces and shards could be caught rather than having them contaminating the larger sample space of the MPMS unit.

 For the study of the mechanical properties of the crystals, a micropillar was fabricated on a sample using Ga+ ion focused-ion beam (Helios 460F1, Thermo Fisher Scientific). The sample was attached to an aluminum stage using 821-3 Crystalbond™ 509 Clear \cite{TedPella}. Two consecutive nanomechanical compression tests were performed in air at room temperature using a Nanoindenter (iNano, KLA) with a constant displacement rate of 10~nm/s. Then the stage with the sample was heated (ex-situ) on a hot plate, taking $\sim$5 minutes to get from room temperature to 100$^{\circ}$C, and then holding at that temperature for 1 minute. Finally, two more compression tests were performed in air and at room temperature after having heated the sample.

\section{Results}

\begin{figure}[H]
 \centering
 \includegraphics[width=\linewidth]{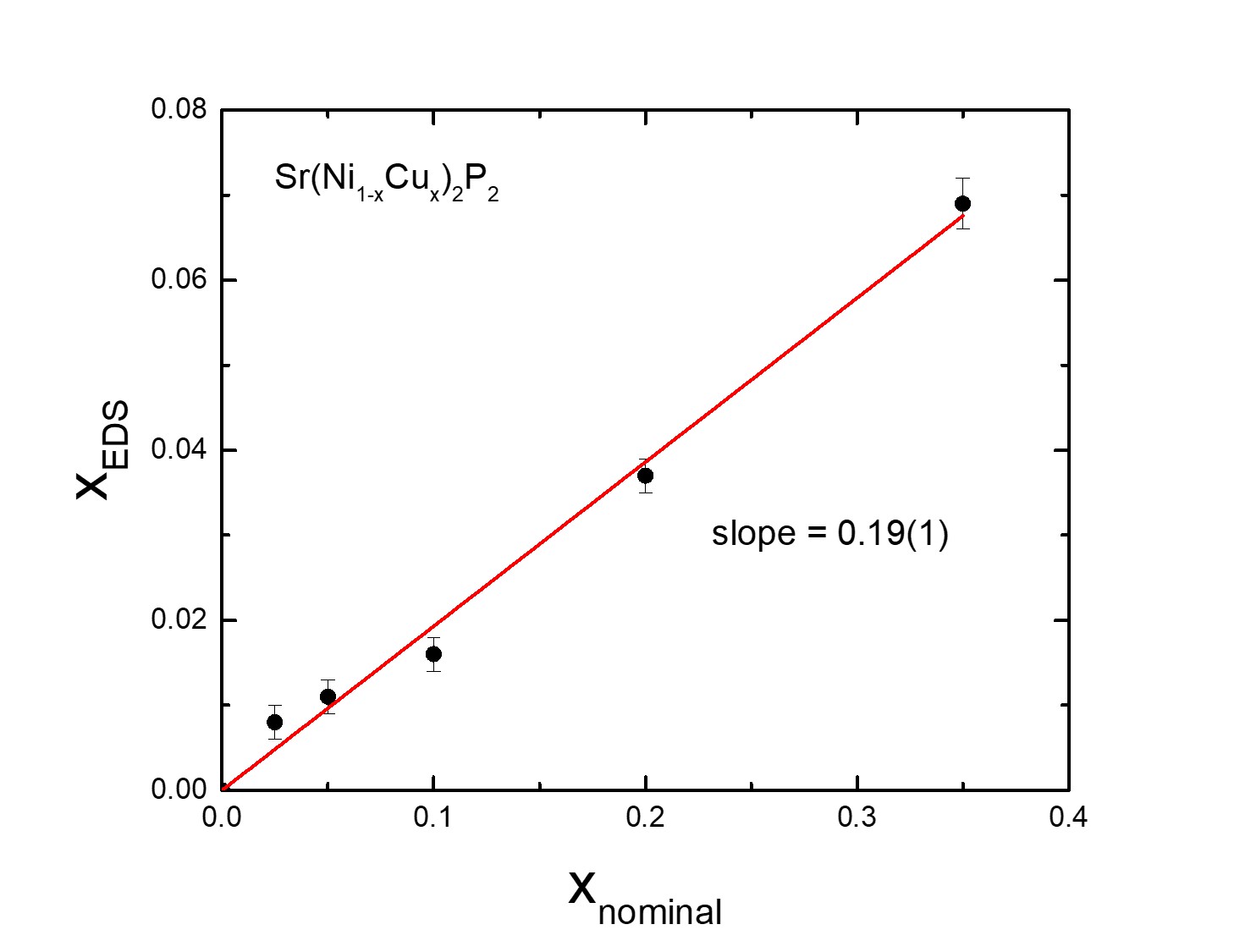}
 \caption{{Fraction of Cu in Sr(Ni$_{1-x}$Cu$_x$)$_2$P$_2$ determined by EDS, $x_{\text{EDS}}$, as a function of the nominal Cu fraction that was originally present in the high-temperature solution that the crystals were grown from. The red line represents the linear fit, for which a zero intercept was forced.}}
 \label{fig:EDS_SrNiCuP}
\end{figure}

Figure \ref{fig:EDS_SrNiCuP} shows the atomic fraction of Cu quantified by EDS, $x_{\text{EDS}}$, as a function of the nominal atomic fraction of Cu, $x_{\text{nominal}}$, in the high-temperature solution from which the crystals were grown. The dependence is approximately linear with a slope of 0.19(1) as indicated by the linear fit shown with a red line, indicating that the crystals grow with less Cu than what is available in the solution. This reduced Cu fraction is uniform across each crystal. The plotted error bars account for the standard deviation of the Cu fraction measured at least four different positions on the crystal face (perpendicular to $c$), as well as different points across a polished edge of the crystal (along $c$), reflecting that the composition of each crystal is relatively homogeneous. From this point onward, $x_{EDS}$ values are simply referred to as $x$, for simplicity.

\begin{figure}[H]
 \centering
 \includegraphics[width=\linewidth]{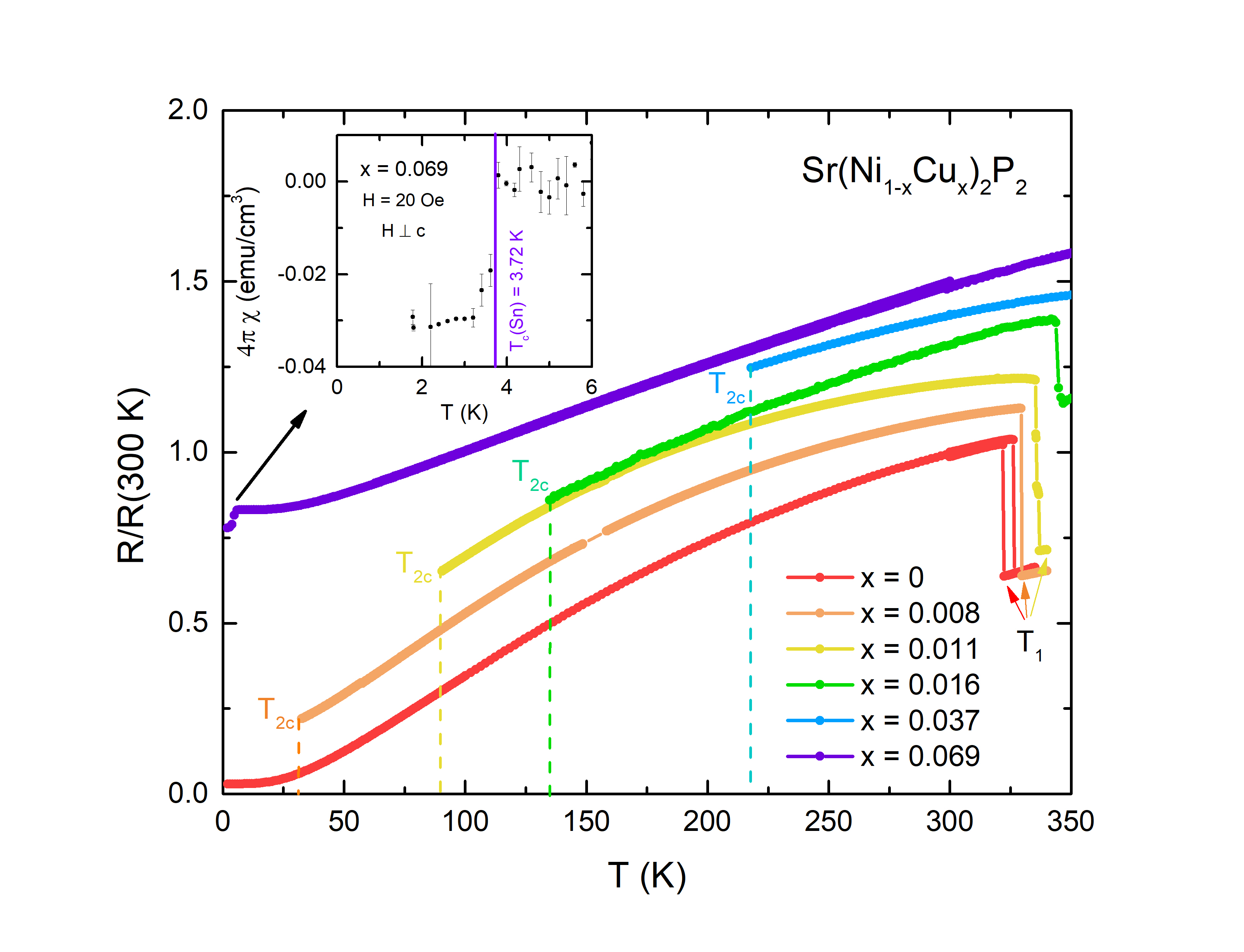}
 \caption{{Main panel: temperature-dependent resistance normalized by its value at 300 K measured upon cooling for Sr(Ni$_{1-x}$Cu$_x$)$_2$P$_2$; a constant offset of 0.1 was added between each curve in order to visualize them more clearly. Inset: magnetic susceptibility measured under a ZFC protocol with a field of 20 Oe applied perpendicular to $c$ for the $x=$0.069, in order to estimate a superconducting volume fraction of less than 0.04. The temperature history protocol in which the datasets were taken is discussed in the main text.}}
 \label{fig:RT_SrNiCuP}
\end{figure}

Temperature dependent resistance data were collected upon cooling for the Sr(Ni$_{1-x}$Cu$_x$)$_2$P$_2$ samples with different $x$, shown in Fig. \ref{fig:RT_SrNiCuP}. The results were normalized by their value at 300 K, and an offset of 0.1 between successive curves was added to facilitate comparison. A sharp steplike feature at temperature $T_1$ can be identified for the samples with $x=0$, 0.008, 0.011,  and 0.016 associated with the tcO$\leftrightarrow$ucT structural transition. For samples with $0.008\leq x\leq 0.037$ the resistance could only be measured down to a temperature $T_{2c}$, since the signal was lost due to cracking along the length and width of the bar-shaped samples and/or loss of the contacts. This is similar to what was observed in Ca(Fe$_{1-x}$Co$_x$)$_2$As$_2$ at the temperatures corresponding to the ucT$\rightarrow$cT transition, due to the internal stresses induced by this first-order structural phase transformation \cite{Ran2012}. This suggests that a similar thing may be occurring in Sr(Ni$_{1-x}$Cu$_x$)$_2$P$_2$, but corresponding to the tcO$\rightarrow$cT transition.

The samples were measured upon cooling from 350 K, in order to capture the step corresponding to $T_1$, if present. The samples with $x=0$ and 0.069 did not break at any temperature, so the resistance could also be measured upon warming. The sample with $x=0$ reveals a small hysteresis at $T_1$ with a width of $5 K$ associated with the tcO$\leftrightarrow$ucT. The low temperature feature observed for the sample with $x=0.069$ is likely due to a small amount of superconducting Sn in the sample, since it was grown out of Sn flux.  As shown in the inset of Fig. \ref{fig:RT_SrNiCuP}, the low-field magnetic zero-field cooled susceptibility is consistent with the superconducting transition of Sn at 3.72 K, corresponding to less than 4\% of the total volume of the sample, which is enough to manifest in the resistance measurements. 

\begin{figure}[H]
 \centering
 \includegraphics[width=\linewidth]{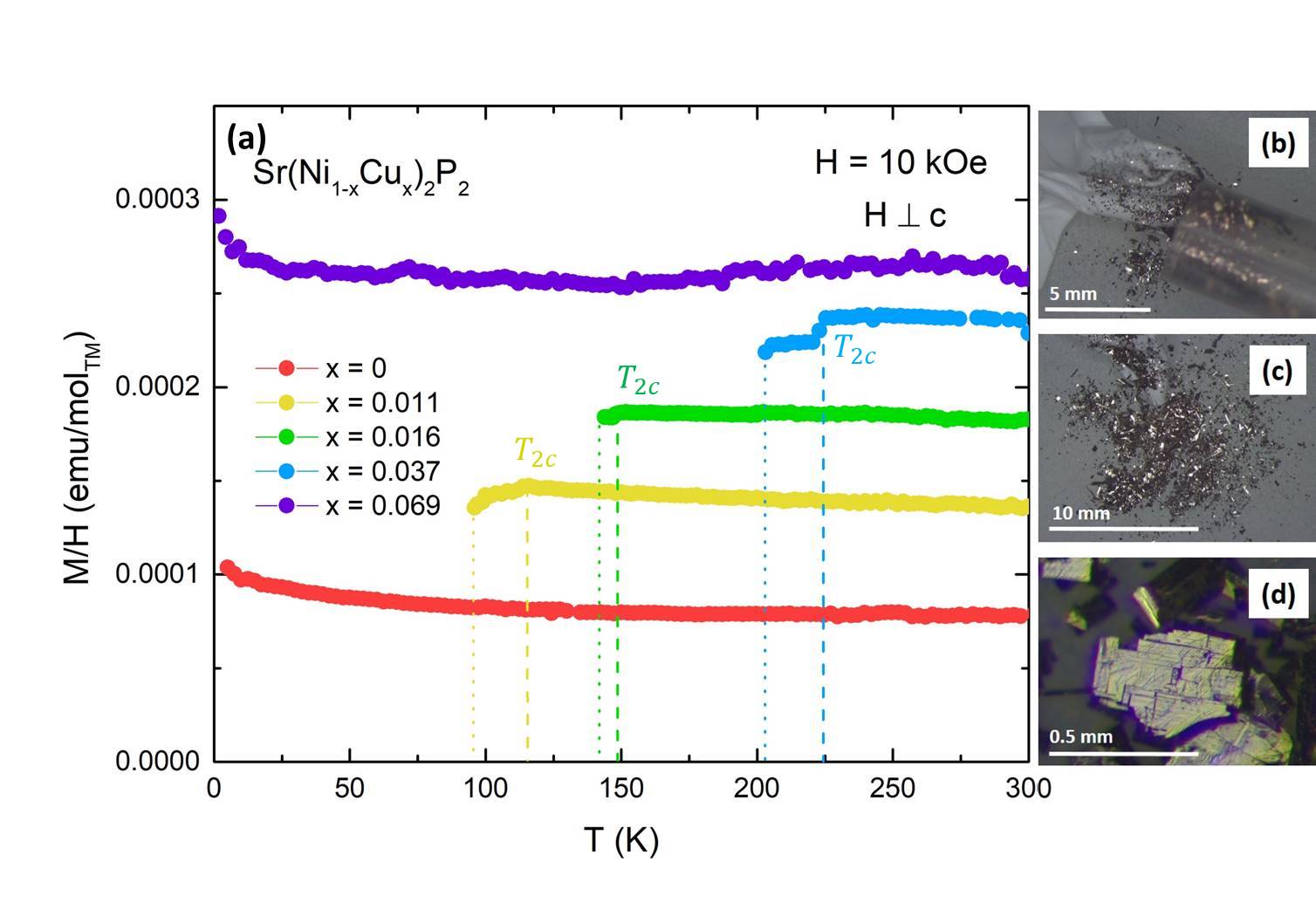}
 \caption{{(a) Temperature-dependent magnetization normalized by the field of 10 kOe applied perpendicular to the crystallographic $c$-axis, measured upon cooling for Sr(Ni$_{1-x}$Cu$_x$)$_2$P$_2$; a constant offset of 0.00005 emu/mol$_{TM}$ was added between each curve in order to visualize them more clearly. The vertical dashed lines indicate the transition temperature $T_{2c}$, and the vertical dotted lines indicate the temperature at which the signal was completely lost due to shattering of the sample. The values of $T_1$ are not observed as they are above the maximum temperature explored in these measurements. (b)-(d) Photos (with increasing magnification) of the shattered crystal fragments for $x=$ 0.016 after cycling through the cT transition.}}
 \label{fig:MT_SrNiCuP}
\end{figure}

A similar signature of $T_{2c}$ is observed in the temperature-dependent magnetization normalized by the field of 10 kOe applied perpendicular to the $c$-axis, shown in Fig. \ref{fig:MT_SrNiCuP}(a). Figures \ref{fig:MT_SrNiCuP}(b) and \ref{fig:MT_SrNiCuP}(c) show the state of a single crystal after being cooled below $T_{2c}$, shattering into pieces that fell to the end of the plastic straw covered with Teflon tape, as they are too small to be held between the two inner straws. A magnified image on one of the pieces is given in Fig. \ref{fig:MT_SrNiCuP}(d), allowing one to appreciate its internal fracturing as well. After the sample pieces fall to the bottom of the straw, there is no proper signal at the original location that allows to continue measuring the sample. In most cases there is a minor step-like feature at a slightly higher temperature, suggesting that part of the sample may have broken and fallen at this temperature, but with most of the sample remaining at its original location. Hence, the temperature $T_{2c}$ was assumed to be the higher temperature step (at which the low-temperature phase most likely nucleated, indicated with a dashed line), and the uncertainty was taken as the distance between this temperature and the temperature at which the signal was lost (indicated with a dotted line).

Both temperatures, $T_1$ and $T_{2c}$, increase with increasing Cu fraction, $x$, although $T_{2c}$ increases much faster. For $x=0.069$, $T_{2c}$ is not present in the measured range and is assumed to be above the highest measured temperature of 350 K. Hence, as opposed to the samples with lower $x$, we were able to measure the resistance and magnetization of the cT state down to 1.8 K. 

\begin{figure}[H]
 \centering
 \includegraphics[width=\linewidth]{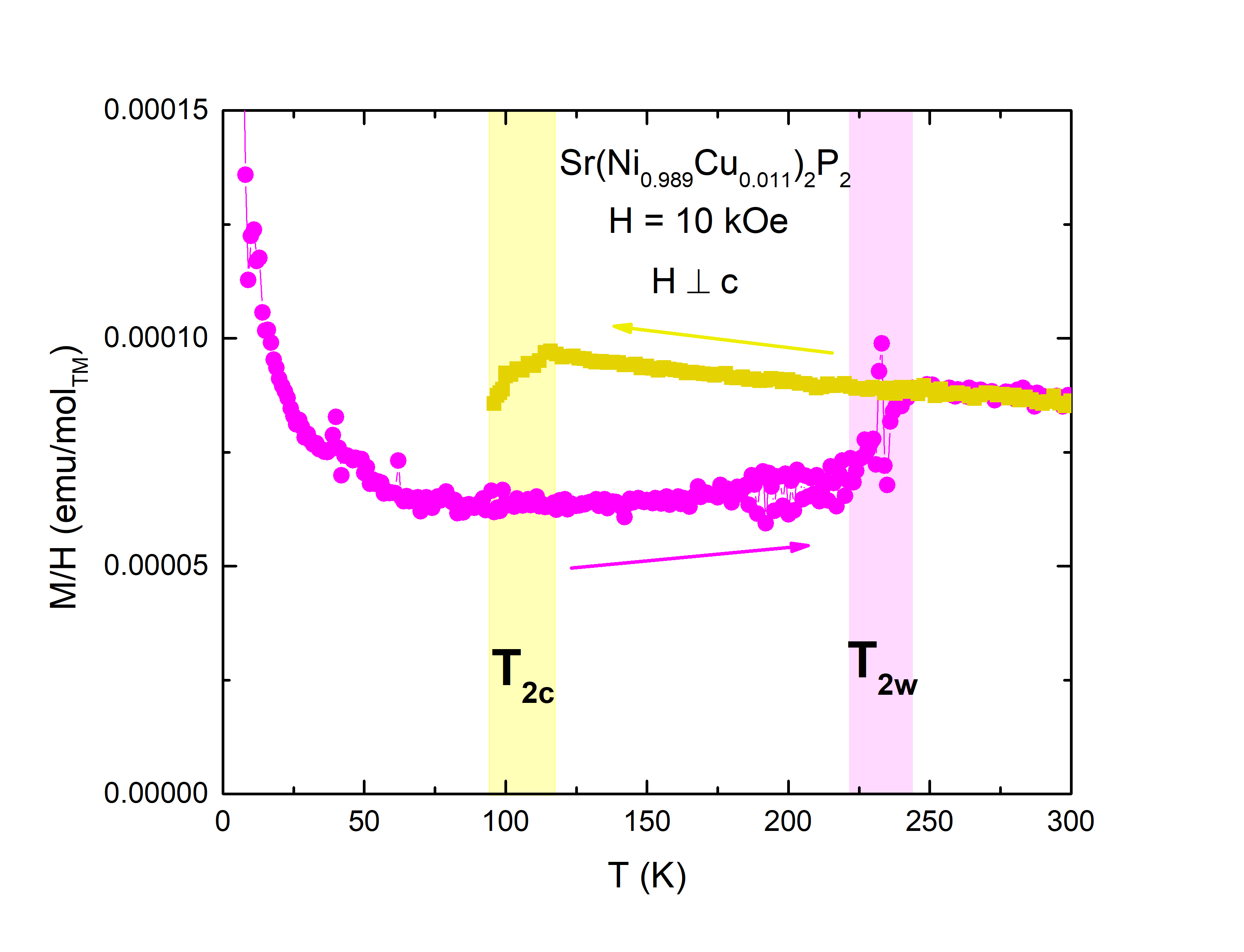}
 \caption{{(a) Temperature-dependent magnetization normalized by the field of 10 kOe applied perpendicular to the crystallographic $c$-axis, measured for Sr(Ni$_{0.989}$Cu$_{0.011}$)$_2$P$_2$ upon cooling (yellow) and warming (magenta) as indicated by the corresponding arrows; the ranges corresponding to the transitions $T_{2w}$ and $T_{2c}$ are shaded in gray. The sample mass was measured before each measurement, in order to be able to properly normalize the two curves.}} 
 \label{fig:MT_SrNiCuP_11pc}
\end{figure}

When measuring the sample with $x=0.011$, the sample was initially cooled to the base temperature before measuring. Fortunately, it was still possible to center and measure, as there was a small portion of the sample that remained in the original location inside the straw after cooling below $T_{2c}$. This allowed for low-temperature measurements upon warming, plotted in magenta in Fig. \ref{fig:MT_SrNiCuP_11pc}, which show a transition at $T_{2w}\sim230$~K. After reaching 300~K, the straw was removed from the MPMS, the remaining piece was carefully removed from the straw, weighed in order to get a proper normalization, and mounted back in the straw. The pieces that had shattered and fell to the end of the straw were removed/cleaned for the following part of the measurement. The straw with the piece that had remained whole was placed back into the MPMS at 300~K and measured upon cooling (plotted in yellow in Fig. \ref{fig:MT_SrNiCuP_11pc}), down to $\sim 100$ K where it finally shattered and all signal was lost. The drop in $\sim 12\%$ of the susceptibility is consistent with that observed in Co- and Rh-substituted SrNi$_2$P$_2$ \cite{Schmidt2023,Schmidt2025}, which are most likely a consequence of changes in the electronic structure that impact on the Pauli/Landau components of susceptibility.

\begin{figure*}
 \centering
 \includegraphics[width=0.9\linewidth]{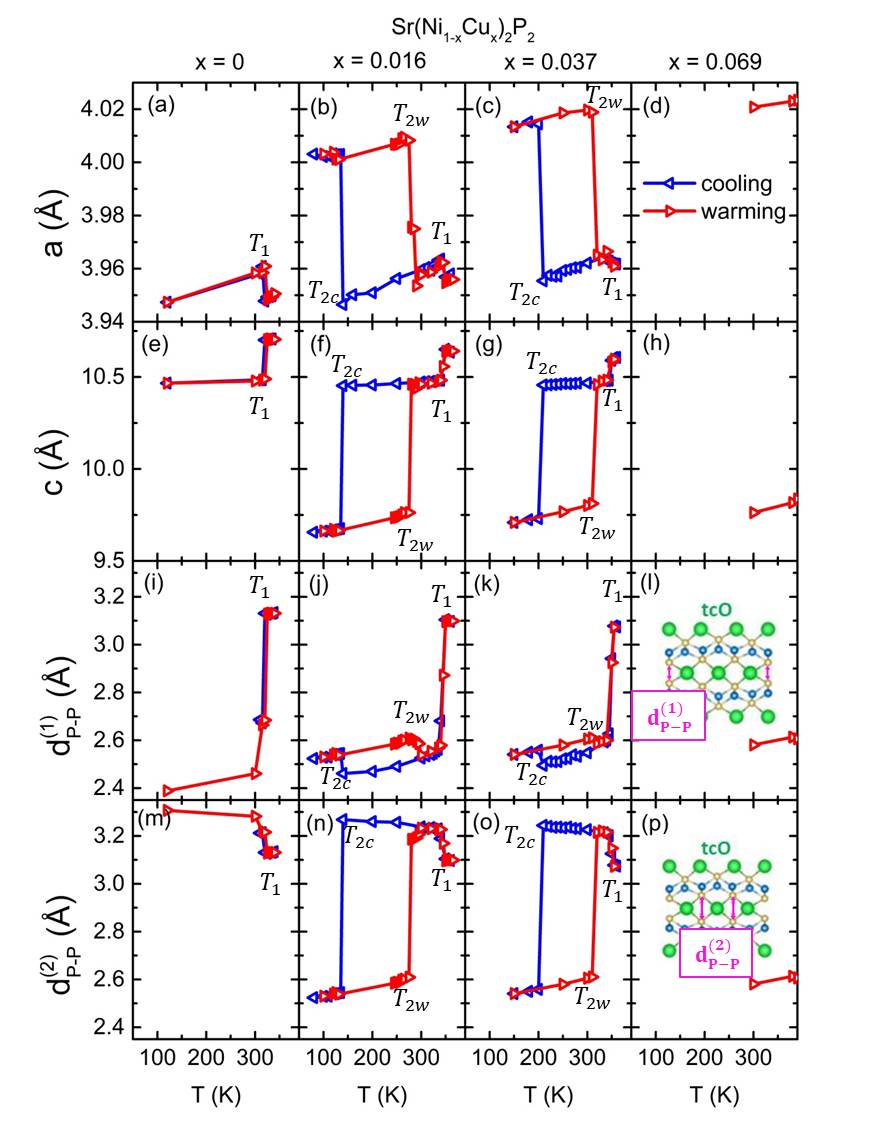}
 \caption{{(a-h) Temperature-dependent lattice parameters and (i-p) P-P distances of Sr(Ni$_{1-x}$Cu$_x$)$_2$P$_2$ with (a,e,i,m) $x=0$, (b,f,j,n) $x=0.016$, (c,g,k,o) $x=0.037$, and (d,h,l,p) $x=0.069$, determined by single-crystal x-ray diffraction; $d_{\text{P-P}}^{(1)}$ and $d_{\text{P-P}}^{(2)}$ correspond to the shorter and longer distance between P atoms across the Sr layers, indicated with magenta arrows in the structure diagrams of panels (l) and (p), respectively. The values of $b$ are not displayed given that they are equal to $a$ for the ucT phase, and to $3a$ for the tcO phase.}}
 \label{fig:scxrd_temp_SrNiCuP}
\end{figure*}

In order to confirm the nature of $T_1$, $T_{2w}$ and $T_{2c}$, we performed temperature-dependent single-crystal x-ray diffraction (XRD) on selected compositions. In particular, the samples with $x=0.016$ and $x=0.037$ were chosen since $T_1$ and $T_{2c}$ lie within the measurement range of the equipment (80 to 400 K). The temperature dependence of the obtained $a$ lattice parameters for the samples with $x=0$, 0.016, 0.037, and 0.069 are shown in Figs. \ref{fig:scxrd_temp_SrNiCuP}(a)-(d); and the dependence of the $c$ lattice parameters are shown in Figs. \ref{fig:scxrd_temp_SrNiCuP}(e)-(h). The values of $b$ are not displayed given that they are equal to $a$ for the ucT phase, and to $3a$ for the tcO phase, consistent with previous reports \cite{Schmidt2025}. A careful refinement of the data allowed us to obtain the position of the P atoms in the unit cell, from which the P-P distances can be determined. In the ucT and cT states, the P-P distances across the Sr layers are all equal, but in the tcO state there are two distinct distances, $d_{\text{P-P}}^{(1)}$ and $d_{\text{P-P}}^{(2)}$ corresponding to the bonded and unbonded pairs, respectively. The shorter distance $d_{\text{P-P}}^{(1)}$ is plotted in Figs. \ref{fig:scxrd_temp_SrNiCuP}(i)-(l) and indicated with magenta arrows in the structure diagram in Fig. \ref{fig:scxrd_temp_SrNiCuP}(l); and the longer distance $d_{\text{P-P}}^{(2)}$ is plotted in  Figs. \ref{fig:scxrd_temp_SrNiCuP}(m)-(p) and indicated with magenta arrows in the diagram in Fig. \ref{fig:scxrd_temp_SrNiCuP}(p). These results were collected upon cooling (in open blue triangles pointing left) and upon warming (in open red triangles pointing right). 

Above the temperature $T_1$, the structure satisfies that $d_{\text{P-P}}^{(2)}=d_{\text{P-P}}^{(1)}$. Upon cooling below the temperature $T_1$, there is a significant decrease in $d_{\text{P-P}}^{(1)}$ and a weaker increase in $d_{\text{P-P}}^{(2)}$, accompanied by a modest decrease of $c$ and a less noticeable increase in $a$. All of this is consistent with the system transitioning from the high-temperature ucT phase into the intermediate-temperature tcO phase, where the $d_{\text{P-P}}^{(1)}$ corresponds to the distance between the bonded P-P pairs, and $d_{\text{P-P}}^{(2)}$ to the unbonded. There is a minor hysteresis of $\sim$5 K between the transition upon cooling and the inverse transition upon warming, which cannot be appreciated in the scales plotted in Fig. \ref{fig:scxrd_temp_SrNiCuP}. These results are consistent with those observed for the same transition that occurs at lower temperatures for Sr(Ni$_{1-x}$Rh$_x$)$_2$P$_2$ \cite{Schmidt2025}. This transition is not detected for the samples with $x=0.069$ shown in Figs. \ref{fig:scxrd_temp_SrNiCuP}(d), \ref{fig:scxrd_temp_SrNiCuP}(h), \ref{fig:scxrd_temp_SrNiCuP}(l), and \ref{fig:scxrd_temp_SrNiCuP}(p). The values of $a$, $c$, $d_{\text{P-P}}^{(1)}$ and $d_{\text{P-P}}^{(2)}$ measured in the range $300\ \text{K}\leq T\leq 390\ \text{K}$ are consistent with the lowest temperature phase observed in $x=0.016$ and $x=0.037$. This indicates that if $T_1$ exists for the sample with $x=0.069$, it is above the highest measured temperature of 390 K.

Upon further cooling below $T_{2c}$, there is a sharp decrease in $d_{\text{P-P}}^{(2)}$ and a smaller increase in $d_{\text{P-P}}^{(1)}$, causing them to become equal. This is accompanied by a large sudden decrease in $c$ and increase in $a$. We can therefore state that at $T_{2c}$ the structure transitions from the intermediate-temperature tcO state to the low temperature cT, where all the P-P pairs are bonded. These results also reveal that the P-P bond length in the cT state is slightly longer than the length of the bonded pairs in the tcO state. However, the $c$ lattice parameter is longer in the tcO state, since two-thirds of the P-P pairs are not bonded across the Sr layers. This transition has a significantly larger thermal hysteresis than the ucT$\leftrightarrow$tcO transition at $T_1$. Upon warming, the structure returns to its tcO state only above a temperature $T_{2w}$ that is more than 100 K higher than $T_{2c}$. The temperature $T_{2w}$ is not observed in the resistance measurements shown in Fig. \ref{fig:RT_SrNiCuP} since it is only observed upon warming, and the resistance measurements could only be carried out for cooling until $T_{2c}$, below which the sample broke. The same is true for most of the magnetization measurements, with the exception of the sample with $x=0.011$, shown in Fig. \ref{fig:MT_SrNiCuP_11pc}.

\begin{figure}[H]
 \centering
 \includegraphics[width=\linewidth]{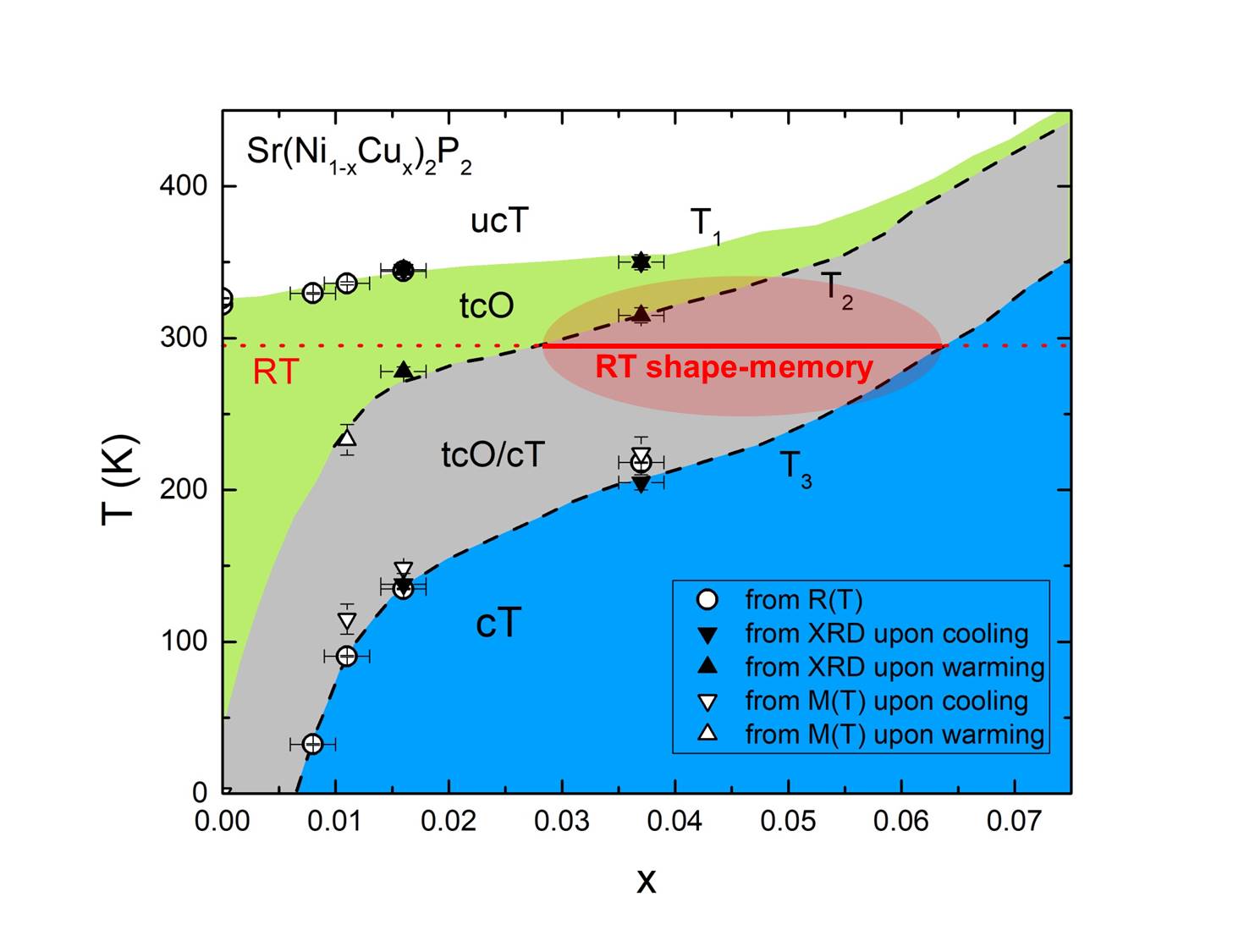}
 \caption{{Temperature-composition phase diagram of Sr(Ni$_{1-x}$Cu$_x$)$_2$P$_2$. The transitions determined from resistance, XRD and magnetization measurements are represented with open circles, solid triangles and solid circles, respectively. The triangles pointing down and up were obtained upon cooling and heating, respectively. The white, green and blue regions correspond to the ucT, tcO and cT phases, respectively. The gray region delimited by the two dashed lines (determined by $T_{2w}$ and $T_{2c}$) corresponds to the hysteresis region, where either tcO or cT can be metastable. The red dotted line indicates the room temperature, and the compositions highlighted with the red oval and solid red line correspond to those that can potentially exhibit room-temperature shape-memory.}}
 \label{fig:phase_diagram_SrNiCuP}
\end{figure}

The phase diagram in Fig. \ref{fig:phase_diagram_SrNiCuP} summarizes the information including all transition temperatures determined by resistance (open circles), single-crystal XRD (black solid triangles), and magnetization (open triangles) measurements. The region where the ucT phase is stable is indicated in white, the region where the cT phase is stable is indicated in blue, and the region where the tcO phase is stable is indicated in green. There is an intermediate region between the green and blue regions, shown in gray, that is delimited by the dashed lines corresponding to $T_{2w}$ and $T_{2c}$ (the width of hysteresis given in Fig. \ref{fig:scxrd_temp_SrNiCuP}), which corresponds to the region where both the tcO and cT phases can be metastable, depending on whether the sample has been warmed or cooled. In particular, for the sample with $x=0.037$, the hysteresis is between $T_{2c}=205$ K and $T_{2w}=315$ K, which means that either the tcO or the cT phase can be metastable at room temperature, depending on the temperature history of the sample. The phase diagram suggests that samples with $0.03\lesssim x\lesssim 0.06$, indicated with a red oval in Fig. \ref{fig:phase_diagram_SrNiCuP}, satisfy the same criterion making them potential room-temperature shape-memory materials.

\begin{figure}[H]
 \centering
 \includegraphics[width=\linewidth]{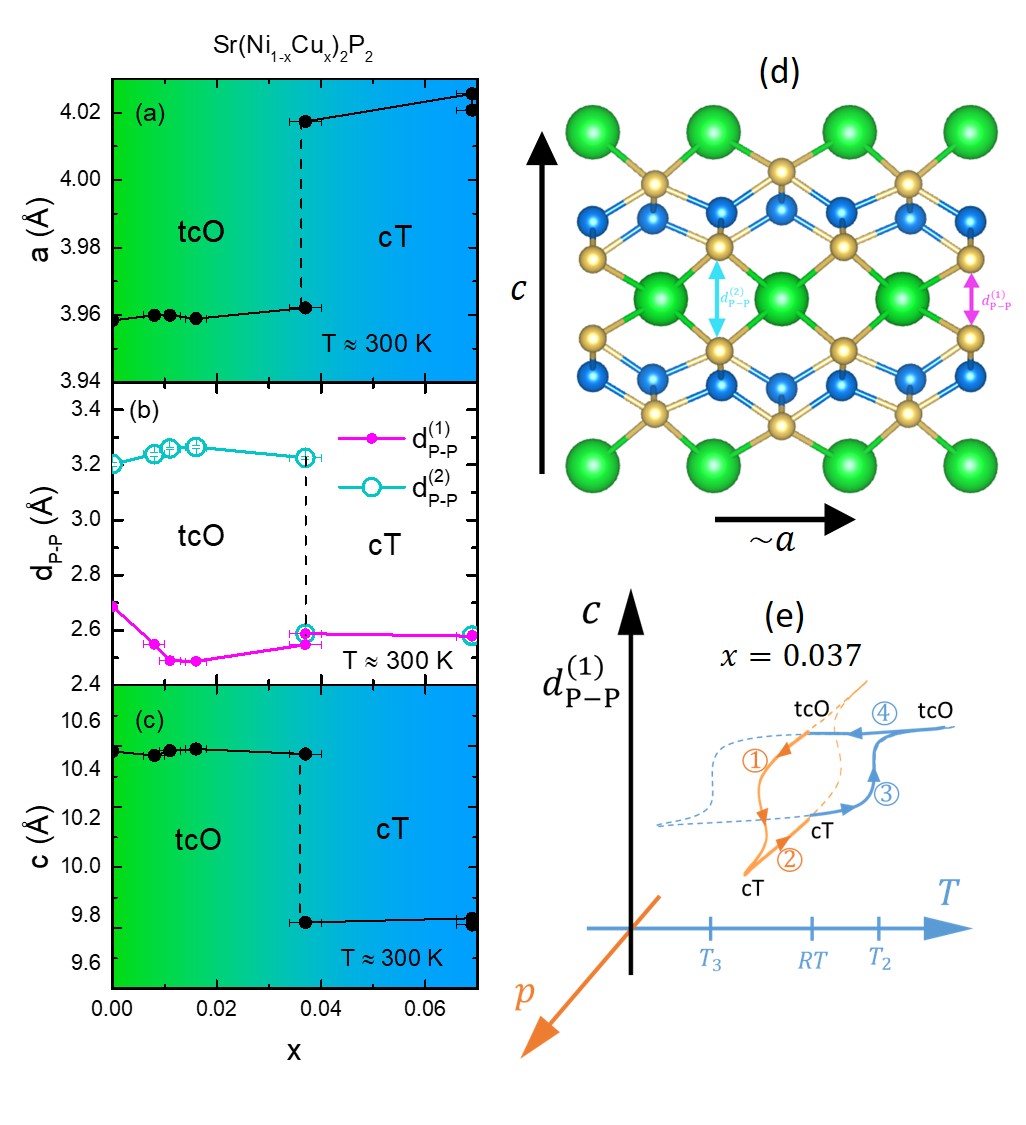}
 \caption{{Dependence of the (a) $a$ and (c) $c$ lattice parameters, as well as (b) the P-P distances across the Sr planes, as a function of the Cu fraction, $x$ in Sr(Ni$_{1-x}$Cu$_x$)$_2$P$_2$. (d) Schematic diagram of the tcO structure indicating $a$, $c$, $d_{\text{P-P}}^{(1)}$ and $d_{\text{P-P}}^{(2)}$. (e) Schematic protocol to illustrate the shape-memory of the sample with $x=0.037$ when combining changes of pressure with changes of temperature.}}
 \label{fig:combined_SrNiCuP}
\end{figure}

Figures \ref{fig:combined_SrNiCuP}(a)-(c) show the room-temperature lattice parameters and P-P distances for all the Sr(Ni$_{1-x}$Cu$_x$)$_2$P$_2$ as a function of the Cu fraction $x$, which are consistent with the tcO phase for $x\leq0.016$ and with the cT phase for $x=0.069$. The possibility of room-temperature metastability of the tcO and cT states in the sample with $x=0.037$ is reflected in Figs. \ref{fig:combined_SrNiCuP}(a)-(c) which show that it has two options for $a$, $c$ and $d_{\text{P-P}}$. 

Previous studies suggest that easily accessible pressures/stresses can induce the tcO$\rightarrow$cT transition. Room-temperature single-crystal x-ray diffraction measurements of SrNi$_2$P$_2$ under hydrostatic pressure show that the system is transformed to the cT state with only 0.4~GPa \cite{Keimes1997}. Since Cu substitution favors P-P bonding, Sr(Ni$_{1-x}$Cu$_x$)$_2$P$_2$ samples with $x<0.069$ should require even less pressure. Mechanical testing studies performed on SrNi$_2$P$_2$ micro-pillars indicate that the cT state is achieved with uniaxial stresses of $\sim0.33$ GPa along the $c$-axis, at room temperature, and that this pressure is reduced to $\sim 0.22$ GPa when cooling the material to 100 K \cite{Xiao2021}. The SrNi$_2$P$_2$ samples have been shown to exhibit room-temperature pseudoelasticity, since they return to the tcO state upon releasing the uniaxial stress.

The possibility of inducing a metastable switching between the tcO and cT phases by applying pressure or strain may be combined with temperature protocols in order to display shape-memory. To illustrate how shape-memory can be achieved, we can follow the protocol schematized with solid curves and arrows in Fig. \ref{fig:combined_SrNiCuP}(e), starting from a sample at room temperature and ambient pressure in the tcO state. In step \textcircled{1}, pressure is applied in order to induce the cT state. In step \textcircled{2}, pressure is released, and the cT phase remains metastable given that the sample is at a temperature between $T_{2c}$ and $T_{2w}$. In step \textcircled{3}, the sample is warmed up above $T_{2w}$, in order to recover the tcO state. Finally, in step \textcircled{4} the sample is allowed to cool back to room temperature, returning to the starting point of the cycle. 

\begin{figure}[H]
 \centering
 \includegraphics[width=\linewidth]{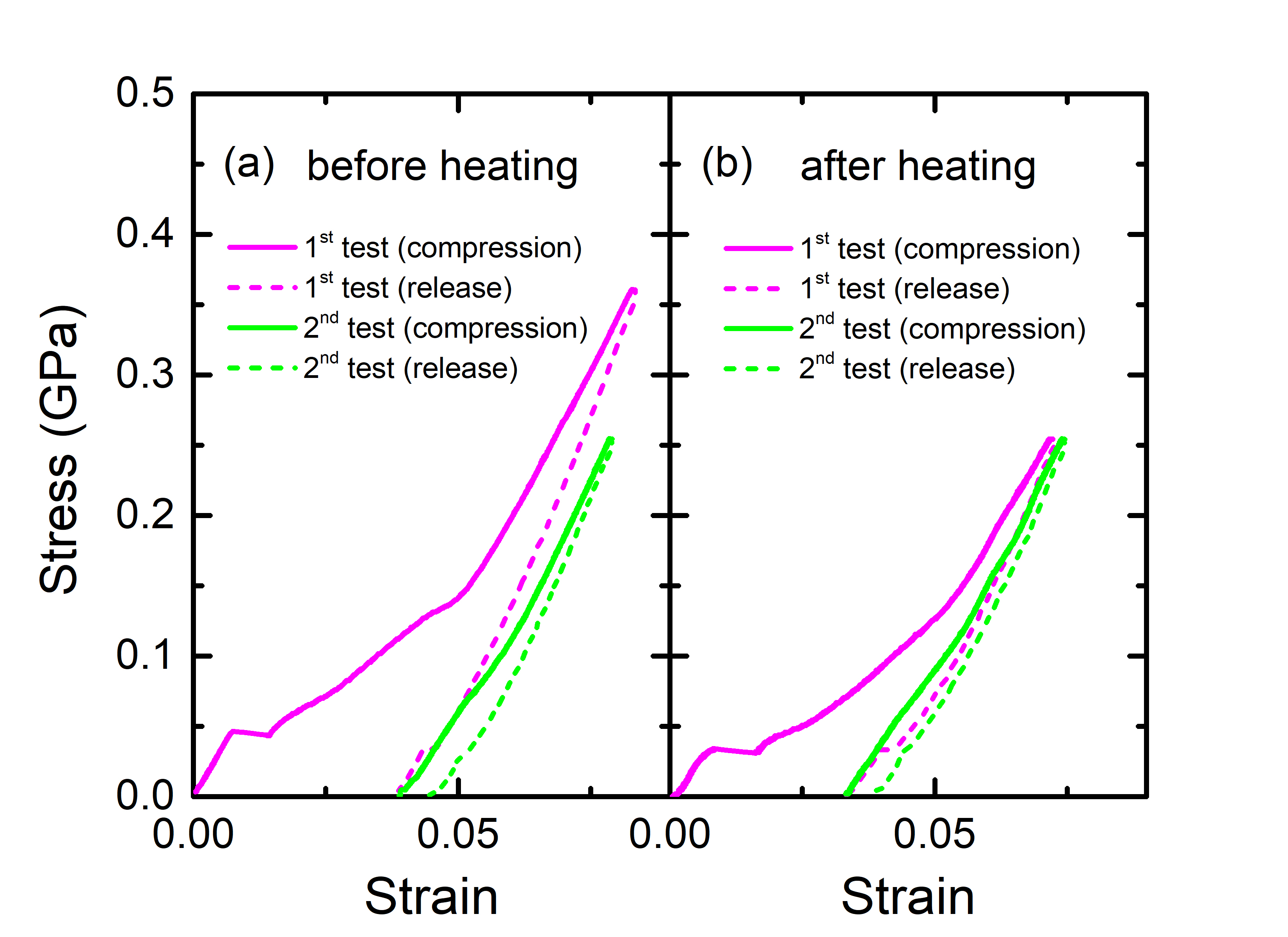}
 \caption{{Uniaxial stress versus strain curves measured along the $c$ axis on single-crystalline micropillars of Sr(Ni$_{0.963}$Cu$_{0.037}$)$_2$P$_2$ at room temperature. (a) Results obtained on the as-grown sample before heating, on the first compression (magenta solid line), first release (magenta dashed line), second compression (green solid line), and second release (green dashed line). (b) Similar results obtained after heating the sample to 373~K and then returning to room temperature.}}
 \label{fig:strain_stress_SrNiCuP}
\end{figure}

This protocol can also be carried out using uniaxial stress rather than hydrostatic pressure. This was tested for a sample with $x=0.037$ and the results are shown in Fig. \ref{fig:strain_stress_SrNiCuP}. A first compression (magenta solid line in Fig. \ref{fig:strain_stress_SrNiCuP}(a)) on an as-grown sample in the tcO state displays a nonlinear behavior (that extends from $\sim0.05$~GPa to $\sim0.14$~GPa) due to the transition to the cT phase. When releasing this stress (magenta dashed line in Fig. \ref{fig:strain_stress_SrNiCuP}(a)) the behavior is closer to linear and retains a remanent strain of $\sim 0.04$. This remanent strain is consistent with the relative $c$-axis difference between the tcO and cT states for this composition shown in Fig. \ref{fig:combined_SrNiCuP}(b). In fact, we can confirm that the sample remains in this metastable state by performing a second compression test (green curves in Fig. \ref{fig:strain_stress_SrNiCuP}(a)) that exhibits a much smaller hysteresis and a similar elastic behavior as the magenta dashed curve. In order to demonstrate the shape-memory effect, this sample was heated to 100$^{\circ}$C for 1 minute, and then subjected to two more compression tests. The magenta solid line in Fig. \ref{fig:strain_stress_SrNiCuP}(b) corresponding to the first compression test after heating resembles the first compression test done before heating, consistent with recovery of the as-grown state. The rest of the curves shown in Fig. \ref{fig:strain_stress_SrNiCuP}(b) demonstrate that the whole sequence can be repeated after the first heating, with a lower hysteresis of the second cycle corresponding to the elastic behavior of the cT phase.

The temperature of the unstrained sample should not be lowered below $T_{2c}$ (which corresponds to 200~K for the sample with $x=0.037$), because large samples shatter when they transition to the cT state upon cooling. This implies that this material cannot be be used for shape-memory applications below that temperature, but still provides a comfortable temperature range for applications near room temperature.

\section{Discussion and Conclusions}

Whereas we have delineated a region of shape memory behavior at room temperature, it is expected that shape memory effects could exist at lower and/or higher temperatures. The sample with $x=0$ does not display the feature associated with the tcO$\leftrightarrow$cT which could indicate that the cT phase is not stable for this composition. We can confidently conclude that there is no $T_1$ down to 1.8 K (the lowest temperature at which resistance was measured) as indicated by the lower dashed line in Fig. \ref{fig:phase_diagram_SrNiCuP}, so virtually no amount of cooling by itself is enough to induce the cT phase at ambient pressure. However, it is possible that it has a finite $T_{2w}>0$ below which the cT phase would be metastable after having induced it in another way. A protocol that combines pressure/stress and temperature could resolve whether this is the case: (1) the sample is cooled down to the lowest temperature allowed by the experimental setup, (2) pressure is applied on the sample to induce the cT phase, (3) the pressure is released back to ambient pressure, and (4) the sample is measured while warming up in order to see if there is any feature associated to $T_{2w}$. In other words, if the SrNi$_2$P$_2$ sample has a finite $T_{2w}$ it should exhibit shape-memory below this temperature. If the cT state can remain metastable at low temperatures after step (3), there should be a transition back to the tcO state at a temperature $T_{2w}(x=0)<T_{2w}(x=0.016)=278(3)$ K. The order of steps (1) and (2) can also be reversed, without impacting the observation of $T_{2w}$. The mechanical study on SrNi$_2$P$_2$ micro-pillars does not report the behavior of the samples at 100 K upon releasing the uniaxial stress \cite{Xiao2021}. If $T_{2w}>100$ K, then the material should remain in the cT after stress is released at this temperature.

For the sample with $x=0.069$, as mentioned above, the lattice parameters and P-P distances are consistent with the cT phase. No transition is observed for $300\ \text{K}\leq T\leq 390\ \text{K}$. This implies that if $T_{2w}$ exists, the sample needs to be heated to temperatures even higher than 390 K in order to induce the tcO state. These measurements, however, do not provide information about $T_{2c}$, which could be below 390 K. In order to observe it, we would need to increase the temperature of the sample high enough to induce the tcO state, and then measure upon cooling to see if there is any feature associated with $T_{2c}$. It should be noted that the samples were originally grown at 900$^{\circ}$C $=$ 1173~K and cooled to room temperature. Hence, if the samples were in the tcO or ucT state at very high temperatures, the $T_{2c}$ at which the cT phase is induced upon cooling should be higher than room temperature. The dashed lines drawn in Fig. \ref{fig:phase_diagram_SrNiCuP} corresponding to $T_{2w}$ and $T_{2c}$ are drawn in a way that $T_{2w}>$ 390 K and $T_{2c}>$ 300 K for $x=0.069$, but their precise value has not been determined. 

\begin{figure}[H]
 \centering
 \includegraphics[width=\linewidth]{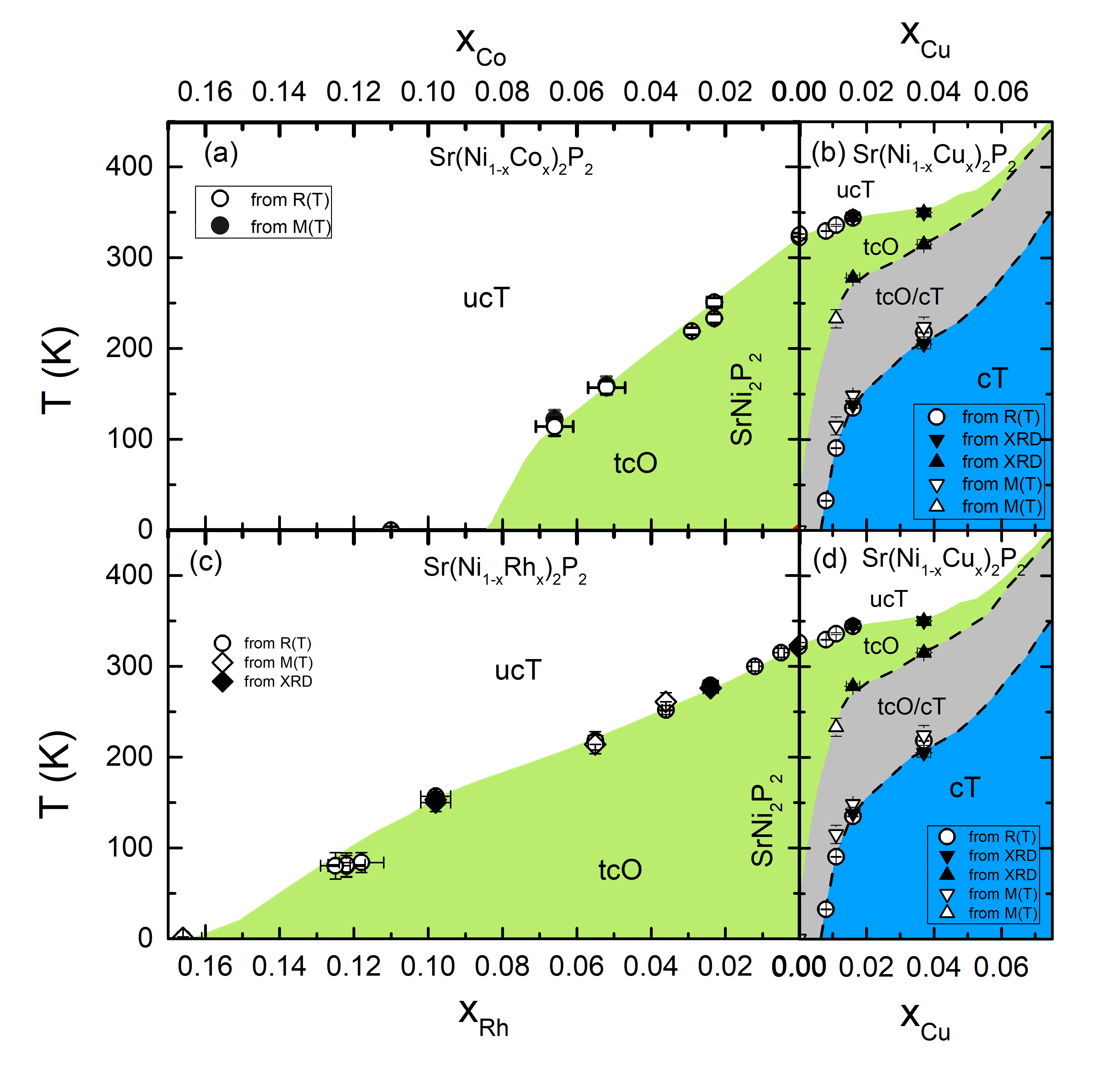}
 \caption{{Combined $T-x$ phase diagram for Sr(Ni$_{1-x}$$M_x$)$_2$P$_2$, with (a) $M=$ Co increasing to the left, (b,d) $M=$ Cu increasing to the right, and (c) $M=$ Rh increasing to the left.}}
 \label{fig:combined_Co_Cu}
\end{figure}

Placing the Cu-substitution in the context of earlier Co- and Rh-substitutions into SrNi$_2$P$_2$ we can see that partial substitution of Cu in Sr(Ni$_{1-x}$Cu$_x$)$_2$P$_2$ promotes P-P bonding across the Sr layers. This manifested as an enhancement of the tcO$\leftrightarrow$ucT transition temperature, as well as the emergence of a new transition at low temperatures into the cT state, evidenced by temperature-dependent resistance and single-crystal XRD measurements. Hence, Cu-substitution exhibits the opposite effect on the structural phase transitions than the previously reported Co- and Rh-substitution series \cite{Schmidt2023,Schmidt2025}. This is represented visually in the combined phase diagrams shown in Fig. \ref{fig:combined_Co_Cu} with the left part of the $x$-axis corresponding to Co (panel a) and Rh (panel c) fractions. 

In contrast with the tcO$\leftrightarrow$ucT transition, the tcO$\leftrightarrow$cT transition has a remarkably large thermal hysteresis across more than 100 K. Both the onset and offset temperatures of this transition increase as the Cu fraction increases, eventually allowing to stabilize the cT state at room temperature for $x\geq 0.037$. In particular, for $x=0.037$ either the cT or the tcO state can be stabilized at room temperature depending on the thermal history of the sample, given that the tcO$\leftrightarrow$cT transition has a hysteresis that spans from 205 to 315 K. 

This suggests that the system can exhibit shape memory at room temperature which, combined with the large maximum recoverable strain and low fatigue of this type of material, makes it a promising candidate for engineering applications. In fact, the maximum recoverable strain in Sr(Ni$_{1-x}$Cu$_x$)$_2$P$_2$ of $\sim14\%$ is significantly larger than the $\sim7\%$ observed in bulk NiTi \cite{Saburi1984}. Additionally, in contrast to NiTi where fatigue is a key issue limiting its applications \cite{Robertson2012}, Sr(Ni$_{1-x}$Cu$_x$)$_2$P$_2$ shows negligible fatigue over 10,000 cycles, due to the fact that shape memory occurs due to the simple breaking and forming of bonds in the strain direction \cite{Xiao2021}. On the other hand, our results have only been tested at a small scale in micropillars fabricated on single crystals, with well controlled uniaxial stress. Further studies should be done on other type of samples such as sintered powder or arc-melted buttons, in order to show that scaling to larger, manufacturable-in-principle quantities is possible, as is done for NiTi.

\section{Acknowledgements}

The authors would like to thank Guilherme Gorgen-Lesseux, Zhuoqi Li and Sushma Kumari for helping with the growth of single crystals used in this work. We appreciate help with the analysis and useful discussions with Rafaela F. S. Penacchio and Tyler J. Slade. Work done at Ames National Laboratory was supported by the U.S. Department of Energy, Office of Basic Energy Science, Division of Materials Sciences and Engineering. Ames National Laboratory is operated for the U.S. Department of Energy by Iowa State University under Contract No. DE-AC02-07CH11358. JS also received support from CONICET during the final writing and analysis stage of this project.

\section{Data Availability}
The data that support the findings of this article are openly
available \cite{data_SrNiCuP}.

\nocite{apsrev41Control}
\bibliographystyle{apsrev4-1}
\bibliography{Introduction_abbreviated.bib}

@CONTROL{apsrev41Control,author="00",editor="1",pages="1",title="0",year="0"}

@article{Newbury2014,
   author = {D. E. Newbury and N. W. M. Ritchie},
   doi = {10.1117/12.2065842},
   issue = {92360H},
   journal = {Proc. SPIE},
   title = {{Rigorous quantitative elemental microanalysis by scanning electron microscopy/energy dispersive x-ray spectrometry (SEM/EDS) with spectrum processing by NIST DTSA-II}},
   volume = {9236},
   pages = {90--106},
   year = {2014},
}

@article{Shatruk2019,
title = {{ThCr$_2$Si$_2$ structure type: The “perovskite” of intermetallics}},
journal = {J. Solid State Chem.},
volume = {272},
pages = {198-209},
year = {2019},
issn = {0022-4596},
doi = {https://doi.org/10.1016/j.jssc.2019.02.012},
url = {https://www.sciencedirect.com/science/article/pii/S0022459619300672},
author = {M. Shatruk}
}

@article{Kreyssig2008,
   author = {A. Kreyssig and M. A. Green and Y. Lee and G. D. Samolyuk and P. Zajdel and J. W. Lynn and S. L. Bud'ko and M. S. Torikachvili and N. Ni and S. Nandi and J. B. Leao and S. J. Poulton and D. N. Argyriou and B. N. Harmon and R. J. McQueeney and P. C. Canfield and A. I. Goldman},
   doi = {10.1103/PhysRevB.78.184517},
   issue = {18},
   journal = {Phys. Rev. B},
   month = {11},
   pages = {184517},
   title = {{Pressure-induced volume-collapsed tetragonal phase of CaFe$_2$As$_2$ as seen via neutron scattering}},
   volume = {78},
   year = {2008},
}

@article{Ran2012,
  title = {{Control of magnetic, nonmagnetic, and superconducting states in annealed Ca(Fe${}_{1\ensuremath{-}x}$Co${}_{x}$)${}_{2}$As${}_{2}$}},
  author = {Ran, S. and Bud'ko, S. L. and Straszheim, W. E. and Soh, J. and Kim, M. G. and Kreyssig, A. and Goldman, A. I. and Canfield, P. C.},
  journal = {Phys. Rev. B},
  volume = {85},
  issue = {22},
  pages = {224528},
  numpages = {16},
  year = {2012},
  month = {Jun},
  publisher = {American Physical Society},
  doi = {10.1103/PhysRevB.85.224528},
  url = {https://link.aps.org/doi/10.1103/PhysRevB.85.224528}
}

@article{Keimes1997,
author = {Keimes, V. and Johrendt, D. and Mewis, A. and Huhnt, C. and Schlabitz, W.},
title = {{Zur Polymorphie von SrNi$_2$P$_2$ sowie zur Kristallstruktur von BaNi$_2$P$_2$}},
journal = {Z. Anorg. Allg. Chem.},
volume = {623},
number = {11},
pages = {1699-1704},
doi = {https://doi.org/10.1002/zaac.19976231104},
year = {1997}
}

@article{Kaluarachchi2017,
   author = {U. S Kaluarachchi and V. Taufour and A. Sapkota and V. Borisov and T. Kong and W. R. Meier and K. Kothapalli and B. G Ueland and A. Kreyssig and R. Valenti and R. J. McQueeney and A. I. Goldman and S. L. Bud'ko and P. C. Canfield},
   doi = {10.1103/PhysRevB.96.140501},
   issue = {14},
   journal = {Phys. Rev. B},
   month = {10},
   pages = {140501(R)},
   title = {{Pressure-induced half-collapsed-tetragonal phase in CaKFe$_4$As$_4$}},
   volume = {96},
   year = {2017},
}

@article{Xiao2021,
   author = {S. Xiao and V. Borisov and G. Gorgen-Lesseux and S. Rommel and G. Song and J. M. Maita and M. Aindow and R. Valentí and P. C. Canfield and S.-W. Lee},
   doi = {10.1021/acs.nanolett.1c01750},
   issue = {19},
   journal = {Nano Lett.},
   month = {10},
   pages = {7913-7920},
   title = {{Pseudoelasticity of SrNi$_2$P$_2$ Micropillar via Double Lattice Collapse and Expansion}},
   volume = {21},
   year = {2021},
}

@article{Canfield2020,
   title={New materials physics},
  author={Canfield, P. C},
  journal = {Rep. Prog. Phys},
  volume={83},
  number={1},
  pages={016501},
  year={2019},
  publisher={IOP Publishing}
}

@article{CanfieldP.C.KongT.KaluarachchiU.S.2016,
   author = {N. H. Jo and Canfield, P. C. and Kong, T. and Kaluarachchi, U. S.},
   issue = {1},
   journal = {Philos. Mag.},
   pages = {84-92},
   title = {{Use of frit-disc crucibles for routine and exploratory solution growth of single crystalline samples}},
   volume = {96},
   year = {2016},
}

@article{Yu2009,
   author = {W Yu and A A Aczel and T J Williams and S. L. Bud'ko and N Ni and P C Canfield and G M Luke},
   doi = {10.1103/PhysRevB.79.020511},
   issue = {2},
   journal = {Phys. Rev. B},
   month = {1},
   pages = {020511(R)},
   title = {{Absence of superconductivity in single-phase CaFe$_2$As$_2$ under hydrostatic pressure}},
   volume = {79},
   year = {2009},
}

@article{Song2019,
   author = {G. Song and V. Borisov and W. R. Meier and M. Xu and K. J. Dusoe and J. T. Sypek and R. Valentí and P. C. Canfield and S.-W. Lee},
   doi = {10.1063/1.5087279},
   issue = {6},
   journal = {APL Mater.},
   title = {{Ultrahigh elastically compressible and strain-engineerable intermetallic compounds under uniaxial mechanical loading}},
   volume = {7},
   year = {2019},
}

@article{Bakst2018,
   author = {C. R. Weinberger and I. N. Bakst and K. J. Dusoe and G. Drachuck and J. R. Neilson and P. C. Canfield and S. -W. Lee},
   journal = {Acta Mater},
   pages = {224-234},
   title = {{Effects of point defects on the mechanical response of LaRu$_2$P$_2$}},
   volume = {160},
   year = {2018},
}

@article{Gati2012,
   title={{Hydrostatic-pressure tuning of magnetic, nonmagnetic, and superconducting states in annealed Ca(Fe$_{1-x}$Co$_x$)$_2$As$_2$}},
  author={Gati, E. and Kohler, S. and Guterding, D. and Wolf, B. and Knoner, S. and Ran, S. and Bud'ko, S. L. and Canfield, P. C. and Lang, M.},
  journal = {Phys. Rev. B},
  volume={86},
  number={22},
  pages={220511(R)},
  year={2012},
  publisher={APS}
}

@article{Borisov2018,
   author = {V. Borisov and P. C. Canfield and R. Valenti},
   issue = {6},
   journal = {Phys. Rev. B},
   month = {8},
   pages = {064104},
   publisher = {American Physical Society},
   volume = {98},
   year = {2018},
   title = {{Trends in pressure-induced layer-selective half-collapsed tetragonal phases in the iron-based superconductor family $AeA$Fe$_4$As$_4$}}
}

@misc{LSPCeramics, 
author = {{LSP Industrial Ceramics}},
howpublished = {https://lspceramics.com/canfield-crucible-sets-2/}
}

@article{Huyan2025,
  title = {{Near-room-temperature ferromagnetic ordering in the pressure-induced collapsed-tetragonal phase in $\mathrm{SrC}{\mathrm{o}}_{2}{\mathrm{P}}_{2}$}},
  author = {Huyan, S. and Schmidt, J. and Valadkhani, A. and Wang, H. and Li, Z. and Sapkota, A. and Petri, J. L. and Slade, T. J. and Ribeiro, R. A. and Bi, W. and Xie, W. and Mazin, I. I. and Valenti, R. and Bud'ko, S. L. and Canfield, P. C.},
  journal = {Phys. Rev. B},
  volume = {112},
  issue = {4},
  pages = {L041102},
  numpages = {7},
  year = {2025},
  month = {Jul},
  publisher = {American Physical Society},
  doi = {10.1103/vl33-53qm}
}

@article{Juan2009,
  author = {Juan, J. and Nó, M. and Schuh, C},
  journal = {Nat. Nanotechnol.},
  volume = {4},
  issue = {7},
  pages = {415-419},
  numpages = {4},
  year = {2009},
  month = {Dec},
  doi = {10.1038/nnano.2009.142},
  title = {{Nanoscale shape-memory alloys for ultrahigh mechanical damping}}
}

@article{Huber1997,
  author = {Huber, J. E. and Fleck, N. A. and Ashby, M. F.},
  journal = {Proc. R. Soc. Lond. A.},
  volume = {453},
  pages = {2185–2205},
  year = {1997},
  doi = {10.1098/rspa.1997.0117},
  title = {{The selection of mechanical actuators based on performance indices}}
}

@article{Schmidt2023,
  title = {{Effects of Co substitution on the structural and magnetic properties of $\mathrm{Sr}{({\mathrm{Ni}}_{1\ensuremath{-}x}{\mathrm{Co}}_{x})}_{2}{\mathrm{P}}_{2}$}},
  author = {Schmidt, J. and Gorgen-Lesseux, G. and Ribeiro, R. A. and Bud'ko, S. L. and Canfield, P. C.},
  journal = {Phys. Rev. B},
  volume = {108},
  issue = {17},
  pages = {174415},
  numpages = {16},
  year = {2023},
  month = {Nov},
  publisher = {American Physical Society},
  doi = {10.1103/PhysRevB.108.174415},
}

@article{Torikachvili2008,
  title = {{Pressure Induced Superconductivity in ${\mathrm{CaFe}}_{2}{\mathrm{As}}_{2}$}},
  author = {Torikachvili, M. S. and Bud'ko, S. L. and Ni, N. and Canfield, P. C.},
  journal = {Phys. Rev. Lett.},
  volume = {101},
  issue = {5},
  pages = {057006},
  numpages = {4},
  year = {2008},
  month = {Jul},
  publisher = {American Physical Society},
  doi = {10.1103/PhysRevLett.101.057006},
}

@article{Xiang2022,
  title = {{Pressure-temperature phase diagram of $\mathrm{CaK}({\mathrm{Fe}}_{1\ensuremath{-}x}{\mathrm{Mn}}_{x}{)}_{4}{\mathrm{As}}_{4}(x\phantom{\rule{0.16em}{0ex}}=\phantom{\rule{0.16em}{0ex}}0.024)$}},
  author = {Xiang, L. and Xu, M. and Bud'ko, S. L. and Canfield, P. C.},
  journal = {Phys. Rev. B},
  volume = {106},
  issue = {13},
  pages = {134505},
  numpages = {6},
  year = {2022},
  month = {Oct},
  publisher = {American Physical Society},
  doi = {10.1103/PhysRevB.106.134505},
}

@article{Xiang2018b,
  title = {{Pressure-temperature phase diagrams of $\text{CaK}({\mathrm{Fe}}_{1\ensuremath{-}x}{\mathrm{Ni}}_{x}{)}_{4}{\mathrm{As}}_{4}$ superconductors}},
  author = {Xiang, L. and Meier, W. R. and Xu, M. and Kaluarachchi, U. S. and Bud'ko, S. L. and Canfield, P. C.},
  journal = {Phys. Rev. B},
  volume = {97},
  issue = {17},
  pages = {174517},
  numpages = {8},
  year = {2018},
  month = {May},
  publisher = {American Physical Society},
  doi = {10.1103/PhysRevB.97.174517},
  url = {https://link.aps.org/doi/10.1103/PhysRevB.97.174517}
}

@article{Stillwell2019,
  title = {{Observation of two collapsed phases in $\mathrm{CaRbF}{\mathrm{e}}_{4}\mathrm{A}{\mathrm{s}}_{4}$}},
  author = {Stillwell, R. L. and Wang, X. and Wang, L. and Campbell, D. J. and Paglione, J. and Weir, S. T. and Vohra, Y. K. and Jeffries, J. R.},
  journal = {Phys. Rev. B},
  volume = {100},
  issue = {4},
  pages = {045152},
  numpages = {8},
  year = {2019},
  month = {Jul},
  publisher = {American Physical Society},
  doi = {10.1103/PhysRevB.100.045152},
}

@article{Wang2023,
  title = {{Pressure-induced dependence of transport properties and lattice stability of $\mathrm{CaK}{({\mathrm{Fe}}_{1-x}{\mathrm{Ni}}_{x})}_{4}{\mathrm{As}}_{4}$ ($x=0.04$ and 0) superconductors with and without a spin-vortex crystal state}},
  author = {Wang, P. and Liu, C. and Yang, R. and Cai, S. and Xie, T. and Guo, J. and Zhao, J. and Han, J. and Long, S. and Zhou, Y. and Li, Y. and Li, X. and Luo, H. and Li, S. and Wu, Q. and Qiu, X. and Xiang, T. and Sun, L.},
  journal = {Phys. Rev. B},
  volume = {108},
  issue = {5},
  pages = {054415},
  numpages = {7},
  year = {2023},
  month = {Aug},
  publisher = {American Physical Society},
  doi = {10.1103/PhysRevB.108.054415},
  url = {https://link.aps.org/doi/10.1103/PhysRevB.108.054415}
}

@article{Sypek2017,
author = {J. T. Sypek and H. Yu and K. J. Dusoe and G. Drachuck and H. Patel and A. M. Giroux and A. I. Goldman and A. Kreyssig and P. C. Canfield and S. L. Bud’ko and C. R. Weinberger and S.-W. Lee},
title = {{Superelasticity and cryogenic linear shape memory effects of CaFe$_2$As$_2$}},
journal = {Nat. Commun.},
volume = {8},
issue = {1},
pages = {1083},
year = {2017},
doi = {10.1038/s41467-017-01275-z}
}

@article{Omori2013,
  title={{Alloys with long memories}},
  author={Omori, T. and Kainuma, R.},
  journal = {Nature},
  volume={502},
  number={7469},
  pages={42-44},
  year={2013},
  publisher={Nature Publishing Group UK London}
}

@article{Finlayson2023,
  title={{The Contributions by Kazuhiro Otsuka to “Shape Memory and Superelasticity”: A Review}},
  author={Finlayson, T. R.},
  journal = {Shape Mem. Superelasticity},
  volume={9},
  number={2},
  pages={217-230},
  year={2023},
  publisher={Springer}
}

@article{Robinson1976,
author = {A. L. Robinson },
title = {{Metallurgy: Extraordinary Alloys That Remember Their Past}},
journal = {Science},
volume = {191},
number = {4230},
pages = {934-936},
year = {1976},
doi = {10.1126/science.191.4230.934}
}

@article{Birman1997,
    author = {Birman, V.},
    title = {{Review of Mechanics of Shape Memory Alloy Structures}},
    journal = {Appl. Mech. Rev.},
    volume = {50},
    number = {11},
    pages = {629-645},
    year = {1997},
    month = {11},
    doi = {10.1115/1.3101674}
}

@article{Schmidt2025,
  title = {{Tuning the structure and superconductivity of ${\mathrm{SrNi}}_{2}{\mathrm{P}}_{2}$ by Rh substitution}},
  author = {Schmidt, J. and Sapkota, A. and Mueller, C. L. and Xiao, S. and Huyan, S. and Slade, T. J. and Ribeiro, R. A. and Lee, S.-W. and Bud'ko, S. L. and Canfield, P. C.},
  journal = {Phys. Rev. B},
  volume = {111},
  issue = {5},
  pages = {054102},
  numpages = {15},
  year = {2025},
  month = {Feb},
  publisher = {American Physical Society},
  doi = {10.1103/PhysRevB.111.054102}
}

@article{Song2021,
  title = {{Construction of $A\ensuremath{-}B$ heterolayer intermetallic crystals: Case studies of the 1144-phase TM-phosphides $\mathit{AB}$(TM)${}_{4}{\mathrm{P}}_{4}$ (TM=Fe, Ru, Co, Ni)}},
  author = {Song, B. Q. and Xu, Mingyu and Borisov, Vladislav and Palasyuk, Olena and Wang, C. Z. and Valent\'{\i}, Roser and Canfield, Paul C. and Ho, K. M.},
  journal = {Phys. Rev. Mater.},
  volume = {5},
  issue = {9},
  pages = {094802},
  numpages = {13},
  year = {2021},
  month = {Sep},
  publisher = {American Physical Society},
  doi = {10.1103/PhysRevMaterials.5.094802}
}

@article{Boehmer2017,
  title = {{Effect of Biaxial Strain on the Phase Transitions of $\mathrm{Ca}({\mathrm{Fe}}_{1\ensuremath{-}x}{\mathrm{Co}}_{x}{)}_{2}{\mathrm{As}}_{2}$}},
  author = {B\"ohmer, A. E. and Sapkota, A. and Kreyssig, A. and Bud'ko, S. L. and Drachuck, G. and Saunders, S. M. and Goldman, A. I. and Canfield, P. C.},
  journal = {Phys. Rev. Lett.},
  volume = {118},
  issue = {10},
  pages = {107002},
  numpages = {6},
  year = {2017},
  month = {Mar},
  publisher = {American Physical Society},
  doi = {10.1103/PhysRevLett.118.107002}
}

@article{data_SrNiCuP,
author={J. Schmidt and A. Horvarth and S.-W. Lee and S. L. Bud’ko and P. C. Canfeld},
title={{Dataset - Room-temperature shape-memory effect in Sr(Ni$_{1-x}$Cu$_x$)$_2$P$_2$.}},
      journal={Zenodo},
      year={2026},
doi={10.5281/zenodo.19806037}
}

@misc{TedPella, 
author = {{821-3 Crystalbond™ 509 Clear, Ted Pella, inc.}},
howpublished = {https://tedpella.com/technote\_html/821-1-2-3-4-6-TN.pdf}
}

@article{Saburi1984,
title = {{Deformation behavior of shape memory Ti-Ni alloy crystals}},
journal = {Scripta Metallurgica},
volume = {18},
number = {4},
pages = {363-366},
year = {1984},
issn = {0036-9748},
doi = {https://doi.org/10.1016/0036-9748(84)90453-8},
url = {https://www.sciencedirect.com/science/article/pii/0036974884904538},
author = {T. Saburi and M. Yoshida and S. Nenno}
}

@article{Robertson2012,
author = {S W Robertson and A R Pelton and R O Ritchie},
title = {{Mechanical fatigue and fracture of Nitinol}},
journal = {International Materials Reviews},
volume = {57},
number = {1},
pages = {1--37},
year = {2012},
publisher = {Taylor \& Francis},
doi = {10.1179/1743280411Y.0000000009}
}

\end{document}